\documentclass[aps,showpacs,amsmath,amssymb,twocolumn,nofootinbib]{revtex4-1}

\usepackage{cancel}
\usepackage{cases}
\usepackage{subfigure}
\usepackage{graphicx}
\usepackage{dcolumn}
\usepackage{bm}
\usepackage{color}
\usepackage[colorlinks=true,pdfstartview=FitV,linkcolor=blue,citecolor=blue,urlcolor=blue]{hyperref}
\usepackage[mathlines]{lineno}
\usepackage[dvipsnames]{xcolor}
\usepackage{amsmath}
\usepackage{ytableau}

\everymath{\displaystyle}
\begin{document}

\newcommand{\beq}{\begin{equation}}
\newcommand{\eeq}{\end{equation}}
\newcommand{\beqs}{\begin{eqnarray}}
\newcommand{\eeqs}{\end{eqnarray}}

\title{A Model of Interacting Dark Matter and Dark Radiation for $H_{0}$ and $\sigma_{8}$ Tensions}

\author{Gongjun Choi,$^{1}$}
\thanks{{\color{blue}gongjun.choi@gmail.com}}

\author{Tsutomu T. Yanagida,$^{1,2}$}
\thanks{{\color{blue}tsutomu.tyanagida@ipmu.jp}}

\author{Norimi Yokozaki,$^{3}$}
\thanks{{\color{blue}n.yokozaki@gmail.com}}

\affiliation{$^{1}$ Tsung-Dao Lee 
Institute, Shanghai Jiao Tong University, Shanghai 200240, China}

\affiliation{$^{2}$ Kavli IPMU (WPI), UTIAS, The University of Tokyo,
5-1-5 Kashiwanoha, Kashiwa, Chiba 277-8583, Japan}

\affiliation{$^{3}$ Theory Center, IPNS, KEK, 1-1 Oho, Tsukuba, Ibaraki 305-0801, Japan}
\date{\today}

\begin{abstract}
We present a model describing the dark sector (DS) featured by two interactions remaining efficient until late times in the matter-dominated era after recombination: the interaction among dark radiations (DR), and the interaction between a small fraction of dark matter and dark radiation. The dark sector consists of (1) a dominant component cold collisionless DM (DM1), (2) a sub-dominant cold DM (DM2) and (3) a self-interacting DR. When a sufficient amount of DR is ensured and a few percent of the total DM density is contributed by DM2 interacting with DR, this set-up is known to be able to resolve both the Hubble and $\sigma_{8}$ tension. In light of this, we propose a scenario which is logically natural and has an intriguing theoretical structure with a hidden unbroken gauge group ${\rm SU}(5)_{\rm X}\otimes {\rm U}(1)_{\rm X}$. Our model of the dark sector does not introduce any new scalar field, but contains only massless chiral fermions and gauge fields in the ultraviolet (UV) regime. As such, it introduces a new scale (DM2 mass, $m_{\rm DM2}$) based on the confinement resulting from the strong dynamics of ${\rm SU}(5)_{\rm X}$. Both DM2-DR and DR-DR interactions are attributed to an identical long range interaction of ${\rm U}(1)_{\rm X}$. We show that our model can address the cosmological tensions when it is characterized by $g_{\rm X}=\mathcal{O}(10^{-3})-\mathcal{O}(10^{-2})$, $m_{\rm DM2}=\mathcal{O}(1)-\mathcal{O}(100){\rm GeV}$ and $T_{\rm DS}/T_{\rm SM}\simeq0.3-0.4$ where $g_{\rm X}$ is the gauge coupling of ${\rm U}(1)_{\rm X}$ and $T_{\rm DS}$ ($T_{\rm SM}$) is a temperature of the DS (Standard Model sector). Our model explains candidates of DM2 and DR, and DM1 can be any kind of CDM.
\end{abstract}

\maketitle
\section{Introduction}  
The high status of the concordance $\Lambda$CDM (cosmological constant + cold and collisionless dark matter (CDM)) model in the modern cosmology has been consolidated to date on account of its capability to explain the large scale structure of the universe and its evolution successfully. Yet, there are phenomenological issues with which predictions based on the $\Lambda$CDM model disagree. An example for this concerns the small scale problems such as core-cusp problem~\cite{Moore:1999gc}, missing satellite problem~\cite{Moore:1999nt,Kim:2017iwr} and too-big-to-fail problem~\cite{Boylan_Kolchin_2011}. Although these small scale issues might hint for an alternative idea for DM other than CDM, it is also probable that those are due to our incomplete knowledge for evolution of structures in non-linear regime and baryonic physics. 

Another example can be found in values of physical quantities obtained from a fit to cosmic microwave background (CMB) data based on the $\Lambda$CDM model. These quantities include the expansion rate of the universe today ($H_{0}$) and matter fluctuations on $8h^{-1}{\rm Mpc}$ scale characterized by $S_{8}$\footnote{$S_{8}$ is defined to be $S_{8}\equiv\sigma_{8}(\Omega_{m}/0.3)^{0.5}$ where $\sigma_{8}$ is the amplitude of matter fluctuations at $8h^{-1}{\rm Mpc}$ and $\Omega_{m}$ is the fraction of the total energy density of the universe contributed by DM and baryon.}. Values of the two quantities provided by the fit are at odds with $H_{0}$ from the local measurement ~\cite{Riess:2016jrr,Riess:2018byc,Bonvin:2016crt,Birrer:2018vtm} at more than $4\sigma$ level and $S_{8}$ from weak lensing surveys at $2-3\sigma$ level, respectively~\cite{Heymans:2013fya,Abbott:2017wau,Hikage:2018qbn,Hildebrandt:2018yau}. Differing from the small scale problems, now that it is well justified to apply the cosmological perturbation theory to scales associated with these tensions, it seems somewhat less likely that these tensions are attributable to lack of a proper theory describing evolution of perturbations at large scales. Then it might be suitable and worthwhile to ask what kind of new physics could reconcile the mismatch albeit not necessary, given the possibility that systematic errors are responsible for the problem~\cite{Efstathiou:2013via,Freedman:2017yms,Rameez:2019wdt}. 

Among many resolutions taking advantage of hypothetical new dark sector (DS) physics,\footnote{Regarding the Hubble tension, several resolutions are suggested by hypothesizing a new physics relevant to a variety of dark sector entities including dark matter \cite{Berezhiani:2015yta,Anchordoqui:2015lqa,Chudaykin:2016yfk,Chudaykin:2017ptd,Vattis:2019efj,Pandey:2019plg,Choi:2019jck,Choi:2020tqp,Blinov:2020uvz,Choi:2020udy}, dark energy~\cite{Poulin:2018cxd,Agrawal:2019lmo,Dutta:2018vmq,Kumar:2019wfs,Niedermann:2019olb,Sakstein:2019fmf,DiValentino:2019ffd,Yang:2018euj,DiValentino:2019jae,Vagnozzi:2019ezj,Visinelli:2019qqu,Lin:2019qug,Alestas:2020mvb,Niedermann:2019olb,Niedermann:2020dwg}, dark radiation (DR)~\cite{DiBari:2013dna,Berezhiani:2015yta,Buen-Abad:2015ova,Lesgourgues:2015wza,Raveri:2017jto,Ko:2017uyb,DEramo:2018vss,Ko:2016uft,Kreisch:2019yzn,Alcaniz:2019kah,Ko:2016fcd,Gonzalez:2020fdy,Gu:2020ozv}. For $\sigma_{8}$ tension, some inferences about non-trivial properties of DM~\cite{Buen-Abad:2015ova,Lesgourgues:2015wza,Anchordoqui:2015lqa,Buen-Abad:2017gxg,Chudaykin:2017ptd,Raveri:2017jto,Ko:2017uyb,Ko:2016uft,Ko:2016fcd,Chacko:2016kgg,Abellan:2020pmw,Heimersheim:2020aoc} and DE~\cite{DiValentino:2019ffd,Davari:2019tni,Camera:2017tws} were made in efforts to resolve the problem. For more complete list of existing resolutions, we refer readers to Ref.~\cite{DiValentino:2020zio} and Ref.~\cite{DiValentino:2020vvd} where the recent reviews of the resolutions for the Hubble tension and $\sigma_{8}$ tension can be found respectively.} particularly interesting is to consider the DS featured by two interactions among dark radiations (DR), and between DR and DM. The existence of DR reduces the sound horizon ($r_{s}$) at recombination and thereby induces increase in $H_{0}$. 
Especially when the DR is an interacting species rather than the free-streaming one, a larger amount of DR is allowed for the same fit to CMB power spectrum. This makes interacting DR option better than the free-streaming one in addressing $H_{0}$ problem~\cite{Blinov:2020hmc,Baumann:2015rya}. On the other hand, the scattering between DR and DM can slow down the rate of change in DM density perturbation as compared to collisionless DM case, which can help alleviate the $\sigma_{8}$ tension. In accordance with these observations, DS models featured by interactions among DRs, and between DR and DM were considered in Refs.~\cite{Buen-Abad:2015ova,Lesgourgues:2015wza,Ko:2016fcd,Ko:2016uft,Chacko:2016kgg,Ko:2017uyb,Buen-Abad:2017gxg,Raveri:2017jto} in efforts to resolve both $H_{0}$ and $\sigma_{8}$ problems simultaneously.

In this letter, motivated by the aforementioned appealing capabilities, we propose a  theoretically well-justified UV-complete DS model entailing the two interactions described above. We focus on the scenario given in ~\cite{Chacko:2016kgg} where the whole of DM population consists of the two components. The sub-dominant DM component (DM2) is responsible for $\sim2\%$ of DM abundance and interacts with DR while the dominant one (DM1) remains collisionless and cold. DM1 can be any kind of CDM candidates in particle physics. As will be shown in Sec.~\ref{sec:model}, in our model, the two interactions are identically explained by ${\rm U}(1)_{\rm X}$ gauge interaction in the DS. The noteworthy peculiarity of our model is that (1) it does not introduce any new scalar fields so as to be free of an additional hierarchy problem, (2) its structure envisages a somewhat large degrees of freedom for DR necessitating DS temperature lower than the Standard Model (SM) temperature, and (3) it naturally explains the mass of DM2 candidate much smaller than UV-cutoff of the theory based on a hidden strong dynamics even if the DM2 candidate is a vector-like fermion.


\section{Review of Phenomenology}
\label{sec:review} 
In this section, we make a brief review of the physics underlying a resolution to the Hubble tension and the $\sigma_{8}$ tension which we attend to in this letter by referring to Refs.~\cite{Buen-Abad:2015ova,Lesgourgues:2015wza,Chacko:2016kgg,Buen-Abad:2017gxg,Blinov:2020hmc}. The resolution assumes that the DS contains both of the DM and the DR. The DR is fluid-like species with a hypothesized non-gravitational self-interaction. The presence of a non-gravitational interaction between DM and DR is assumed as well. Each of non-gravitational interactions among DRs themselves and between DM and DR is the essential physics resolving both of the cosmological tensions as discussed below.

\subsection{Interacting DR}
\label{sec:InteractingDR}
In the SM, for the time when the SM sector thermal bath temperature satisfies $T_{\rm SM}\lesssim1{\rm MeV}$, the radiation density is given by
\beq
\rho_{\rm rad}=\rho_{\gamma}\left[1+ \frac{7}{8}\left(\frac{4}{11}\right)^{4/3}N_{\rm eff}\right]\,,
\label{eq:rhorad}
\eeq
where $\rho_\gamma$ is the energy density of photons and the second term in the bracket is the active neutrino contribution. Because of non-instantaneous decoupling of neutrinos, the effective number of neutrino predicted by the SM reads $N_{\rm eff}=3.046$ \cite{Mangano:2005cc}.\footnote{Although we take $N_{\rm eff}=3.046$ as the effective number of neutrino in the SM in this paper, we notice that a recent computation for $N_{\rm eff}=3.044$ taking into account three-flavor neutrino oscillations and QED finite temperature corrections up to the next-to-leading order in the SM was reported in Ref.~\cite{Akita:2020szl}.} Any extra contribution to $N_{\rm eff}$ parametrized by $\Delta N_{\rm eff}$ is regarded as a hint for an extension of the SM. We call any kind of contribution to non-vanishing $\Delta N_{\rm eff}$ dark radiation (DR).

The actual value of $\Delta N_{\rm eff}$ is an important issue, but another question as important as it is whether contribution to $\Delta N_{\rm eff}\neq0$ is of a free-streaming or an interacting kind. As with SM neutrinos, DR could be an early decoupled species, starting its free-streaming at a time deep inside the radiation-dominated era. In contrast, if DR enjoys a strong hidden non-gravitational interaction, it could be still an interacting species even until today. It is also possible that DR decouples at a time between BBN and CMB era so that non-zero $\Delta N_{\rm eff}^{\rm BBN}$ and $\Delta N_{\rm eff}^{\rm CMB}$ are attributed to an identical species of different natures. In connection with this question about DR's nature, especially the interacting fluid-like DR has an interesting implication for the recently raised Hubble tension~\cite{Blinov:2020hmc}. 

The comoving sound horizon ($r_{s}$) and the angular diameter distance to the surface of the last scattering ($D_{A}$) are defined to be 
\beq
r_{s}\equiv\int_{0}^{a_{\rm rec}}{\rm d}a\frac{c_{s}(a)}{a^{2}H(a)}\quad,\quad D_{A}\equiv\int_{a_{\rm rec}}^{1}\frac{{\rm d}a}{a^{2}H(a)}\,,
\label{eq:rsDA}
\eeq
where $c_{s}(a)$ is the sound speed of the photon-baryon plasma and $a_{\rm rec}$ is the scale factor at which the recombination takes place. As can be seen in limits of integrals in Eq.~(\ref{eq:rsDA}), $r_{s}$ ($D_{A}$) reflects an evolution of the expansion rate of the universe, $H(a)$, before (after) the recombination era. Now as an extra radiation energy source during CMB era, the existence of DR ($\Delta N_{\rm eff}\neq0$) reduces $r_{s}$ and thus induces increase in $H_{0}$; The angle subtended by the sound horizon at the recombination time is given by $\theta_{s}=r_{s}/D_{A}$. Given a precisely measured $\theta_{s}$, decrease in $r_{s}$ requires a compensating decrease in $D_{A}$, leading to increase in $H_{0}$. This is why introducing extra radiations can give rise to increase in $H_{0}$ in a rough understanding.

However, of course one cannot arbitrarily increase $\Delta N_{\rm eff}$ to make $H_{0}$ inferred from CMB power spectrum close to the local measurements. As a non-vanishing $\Delta N_{\rm eff}$ increases, there arises more suppression of CMB temperature anisotropy power spectrum at small scales~\cite{Hou:2011ec} (a.k.a Silk damping).\footnote{The physical quantity parametrizing the amount of Silk damping is the angular scale ($\theta_{d}$) of the diffusion length of photons ($r_{d}$). $\theta_{s}$ being very well determined, $D_{A}\propto1/H$ can be inferred from $r_{s}\propto1/H$. Now given $\theta_{d}=r_{d}/D_{A}$ and $r_{d}\propto1/\sqrt{H}$, $\theta_{d}\propto\sqrt{H}$ is derived. Therefore, increasing $\Delta N_{\rm eff}$ induces increase in the amount of Silk damping by increasing the Hubble expansion rate and $\theta_{d}$~\cite{Hou:2011ec}.} In other words, given a fixed CMB power spectrum data, achieving a fit of good quality requires a certain physics compensating the damping at small scales. The advantage of having interacting DR precisely lies at this point: a fixed $\Delta N_{\rm eff}\neq0$ contributed by an interacting DR corresponds to the smaller suppression of CMB power spectrum on small scales than the case with a free-streaming DR.\footnote{Aside from the different amount of the suppression at small scales, the peak location of CMB anisotropy power spectrum can be an very instrumental and useful tool to probe the nature of the DR~\cite{Bashinsky:2003tk,Baumann:2015rya,Choi:2018gho,Chacko:2015noa}. For a fixed value of $N_{\rm eff}=3.046+\Delta N_{\rm eff}$, the case with $\Delta N_{\rm eff}\neq0$ contributed by an extra fluid-like DR brings about change in peak locations toward smaller scales (larger $\ell$), as compared to the case with $\Delta N_{\rm eff}\neq0$ contributed by extra free-streaming DR. This phase shift effect is a unique imprint on CMB power spectrum which cannot be mimicked by varying other parameters, breaking degeneracy between $N_{\rm eff}$ and the Helium mass fraction.} Presence of an interacting DR tends to result in the greater gravitational potential at horizon crossing and in so doing lowers the degree of damping of CMB power spectrum at small scales. Hence, in making a fit of the same level, a larger amount of $\Delta N_{\rm eff}$ is allowed for an interacting DR than the case with a free-streaming DR, which results in a higher inferred $H_{0}$ value~\cite{Blinov:2020hmc}. Motivated by this winning attribute, in this letter, we search for and focus on the DR sector parameter space which can delay decoupling of DR in the model until at least the recombination era. In Ref.~\cite{Blinov:2020hmc}, we refer to the case where the amount of free-streaming radiation is fixed to $N_{\rm eff}=3.046$ (contributed by the SM active neutrino) and the interacting fluid-like DR amount ($N_{\rm fld}$) is varied as a free-parameter. From the analysis of constraining $\Lambda$CDM model parameters and $N_{\rm tot}^{\rm CMB}\equiv3.046+\Delta N_{\rm fld}^{\rm CMB}$ based on the Planck TT, TE, EE + BAO likelihood, we read $\Delta N^{\rm CMB}_{\rm fld}\!\simeq\!0.4$ (95\% C.L.).

\subsection{Interaction between DM and DR}
As was analyzed and pointed out in Ref.~\cite{Chacko:2016kgg}, $\sigma_{8}$ problem can be resolved when (i) DM comprises two cold components (DM1 and DM2 with $\rho_{\rm DM1}>\rho_{\rm DM2}$) and (ii) only the subdominant component DM2 ($2\%$ of total DM population) interacts with DR. To put it another way, leaving the regime with $k<\!\!<k_{\rm eq}\simeq0.01h{\rm Mpc}^{-1}$ almost intact, this set-up renders the matter power spectrum at $k\simeq0.2h{\rm Mpc}^{-1}$ suppressed by $\sim10\%$ as compared to what is expected by $\Lambda$CDM model, which is enough for resolving the $\sigma_{8}$ problem. The intuitive way of understanding this is what follows.

Via the Einstein equation, one can see how the matter density perturbation and perturbations to the space-time metric affect evolution of each other. For convenience of the following discussion, let us introduce the density contrast $\delta\equiv\delta\rho/\bar{\rho}$, divergence of comoving velocity $\theta\equiv\partial^{i}v_{i}=\partial^{i}x_{i}/d\tau$ with $d\tau=a{\rm d}t$ (conformal time), the gravitational potential $\Phi$, and the spatial curvature perturbation $\Psi$. The last two are scalar perturbations to 00 and $ii$ ($i=1,2,3$) components of the metric in the Newtonian gauge.\footnote{The energy-momentum tensor $T_{\mu\nu}$ of a free-streaming species has non-vanishing off-diagonal components $T_{ij}$ ($i,j=1,2,3$) because isotropization stops for the free-streaming species. When $T_{ij}\neq0$ holds at least for a single species, the difference between $\Phi$ and $\Psi$ becomes non-zero~\cite{Ma:1995ey,Bashinsky:2003tk}. This implies that the relation $\Phi=\Psi$ is maintained (breaks down) prior to (after) occurrence of a free-streaming species.} By the conservation of the energy and momentum of each species (DM1, DM2 and DR), in the presence of interaction between DM2 and DR, it turns out that a non-gravitational interaction between DM2 and DR makes $\theta_{\rm DM2}$ grow faster than the case without the interaction, which suppresses growth of $\delta_{\rm DM2}$. This prevents $\delta_{\rm DM2}$ from growing after matter-radiation equality ($a=a_{\rm eq}$) whereas $\delta_{\rm DM1}$ grows monotonically at the time. The dark thermal bath of DM2 and DR undergoes the acoustic oscillation between the time slices corresponding to $a_{\rm eq}$ and $a_{\rm rec}$. $\delta_{\rm DM2}$ does not grow until the recombination era is reached. After the recombination, $\delta_{\rm DM2}$ starts to grow monotonically. In the end, as compared to $\Lambda$CDM case (single collisionless CDM with $N_{\rm eff}=3.046$), this kind of evolution of $\delta_{\rm DM1}$ and $\delta_{\rm DM2}$ results in the smaller $\Phi$ (via 00-component of the Einstein equation) for the modes with $k\gtrsim k_{\rm eq}$, which leads to suppression of the matter power spectrum for $k\gtrsim k_{\rm eq}$. In this way, suppression of the matter fluctuation at $k\simeq0.2h{\rm Mpc}^{-1}$ required for solving $\sigma_{8}$ problem can be induced with the aid of DM2-DR interaction.

Motivated by the advantage of the interaction between DM2 and DR, in this letter, we consider the scenario where $2\%$ of DM population communicates with DR until late times after the recombination. Note that even if theoretically the interaction is efficient today by having the associated momentum transfer rate $\Gamma_{\rm DM2-DR}$ greater than $H_{0}$, since $\rho_{\rm DR}$ becomes negligible in comparison with $\rho_{\rm DM2}$ after recombination, DM2 practically becomes free soon after the recombination. Thus there should not be a concern for too much suppression of the matter power spectrum at large scales. Furthermore, because only the small fraction of DM population is coupled to DR, the model we shall consider in the next section is free of any limitation on a strength of the interaction between DM2 and DR.


\section{Model}
\label{sec:model} 
In the previous section, we made a brief discussion about how the DS of our interest can reconcile discrepancies observed in measurements of $H_{0}$ and $S_{8}$. Now in this section, reminding ourselves of necessities of the DS we desire, we go through a series of steps to build a logically compelling particle physics model realizing the scenario. To this end, we begin with the question about what could be a good candidate for the self-interacting DRs.

Above all, we could ask what would be a spin of the DR, assuming that it is a particle kind. Should it be a spin-zero boson, then apparently it must be a pseudo Nambu-Goldstone boson (pNGB) arising from a spontaneously broken global symmetry. However, there are two caveats for this possibility. Firstly, global symmetries are difficult to be exact symmetry because it is broken by quantum gravity effects~\cite{Banks:2010zn}. Secondly, the coupling of pNGB to others (other pNGB and DM2) would include powers of $T_{\rm DS}/F$ where $T_{\rm DS}$ and $F$ are a DS temperature and a breaking scale of the associated global symmetry respectively. Unless a very small $F$ is assumed, $T_{\rm DS}/F$ factor is expected to be too small at the matter-radiation equality time to make two requisite interactions effective. And so we conclude that pNGB is not a good candidate for DR we desire. 

Next, what about having spin-1/2 fermions responsible for the role of the interacting DR? For this possibility, fermions are expected to be chiral, otherwise (if they are vector-like), their natural mass amounts to a UV cut-off of the theory which we take to be the Planck mass here. Then, what could explain the non-gravitational self-interaction among DRs? For this, there can be two options which are the use of a gauge interaction and an Yukawa interaction. For the later, we need to introduce a scalar as a mediator. However, the natural mass scale of the scalar is again the UV cut-off of the theory, which is disadvantageous for keeping the self-interaction of DR effective at least until the recombination era. Contrary to this, the first option using a gauge interaction finds no problem provided the assumed gauge symmetry is unbroken. Therefore, we take massless chiral fermions and massless gauge boson as self-interacting DR candidates in our model and attribute the self-interaction to the gauge interaction of a unbroken ${\rm U}(1)$. Let us call this hidden Abelian gauge group ${\rm U}(1)_{\rm X}$ from here on.\footnote{In~\cite{Buen-Abad:2015ova}, a non-Abelian gauge theory was considered where massless gauge bosons are identified as interacting DRs. However, for our purpose we need to introduce the DM2, which may be Dirac fermions charged under the non-Abelian gauge group. We may construct a model in a similar way in this paper, introducing another non-Abelian gauge theory with massless fermions charged under both of the non-Abelian gauge groups. In this paper, however, we do not discuss such a model, since it becomes more involved.} The self-interaction rate is expected to be $\Gamma\sim g_{\rm X}^{4}T_{\rm DS}$ and thus $\Gamma>H$ holds as far as $(T_{\rm DS}/T_{\rm SM})^{1/4}\times g_{\rm X}>(T_{\rm SM}/M_{P})^{1/4}$ is satisfied where $g_{\rm X}$, $T_{\rm SM}$ and $M_{P}$ are the gauge coupling of ${\rm U}(1)_{\rm X}$, the SM sector temperature and the reduced Planck mass. Later we shall see $T_{\rm DS}/T_{\rm SM}\sim\mathcal{O}(0.1)$ and thus the inequality is easily satisfied for late time universe. 

The next question is about a number of the chiral fermion DR candidate. In order to make the theory UV-complete, ${\rm U}(1)_{\rm X}$ must be anomaly-free and thus it is mandatory to check the anomaly-free conditions for ${\rm U}(1)_{\rm X}$. We assume that SM particles are singlet under ${\rm U}(1)_{\rm X}$. Then, the only relevant anomalies to be checked are ${\rm U}(1)_{\rm X}^{3}$ and ${\rm U}(1)_{\rm X}\times[{\rm gravity}]^{2}$. Requiring cancellation of each anomaly yields
\beq
\sum_{i=1}^{N_{X}}Q_{i}^{3}=0\quad,\quad\sum_{i=1}^{N_{X}}Q_{i}=0\,,
\label{eq:anomalyfree}
\eeq
where $i$ labels the DR chiral fermions, $Q_{i}$s are their ${\rm U}(1)_{\rm X}$ charges, and $N_{\rm X}$ is an unknown number of chiral fermions which satisfies both equations in Eq.~(\ref{eq:anomalyfree}). Beginning with $N_{\rm X}=2$, we are led to the conclusion that the minimum number of chiral fermions of ${\rm U}(1)_{\rm X}$ gauge theory without any vector-like fermions is $N_{X}=5$.\footnote{For $N_{X}=2,4$, it is unavoidable to have vector-like fermions. For $N_{X}=3$, there is no any common solution to Eq.~(\ref{eq:anomalyfree}) because of the Fermat's theorem.} In fact, this framework for a DS modelling was first suggested in \cite{Nakayama:2011dj,Nakayama:2018yvj} under the name ``Number Theory Dark Matter". Possible solutions to Eq.~(\ref{eq:anomalyfree}) for $N_{X}=5$ were found there. Referring to one of those, we assign ${\rm U}(1)_{\rm X}$ charges to five chiral fermions in our DS as 
\beq
\psi_{-9},\quad\psi_{-5},\quad\psi_{-1},\quad\psi_{7}\quad\psi_{8}\,,
\label{eq:fivechiralfermions}
\eeq
where the subscripts for each field indicate ${\rm U}(1)_{\rm X}$ charges. To wrap it up in a nutshell, from the anomaly free conditions, we found that theoretically well-justified number of fermions for DR in our model is five and we shall take chiral fermions in Eq.~(\ref{eq:fivechiralfermions}) as our DR candidates along with the ${\rm U}(1)_{\rm X}$ gauge boson.

Concerning DM2, the crucial thing to be considered is to guarantee its strong enough non-gravitational interaction with DR. As already pointed out above, since the ${\rm U}(1)_{\rm X}$ gauge theory is strongly-coupled one, we can just have the ${\rm U}(1)_{\rm X}$ gauge interaction responsible also for DM2-DR interaction. Then, what could be a DM2 candidate? Now that DM2 is charged under ${\rm U}(1)_{\rm X}$ which should be unbroken, it is better to identify DM2 with a fermion charged under ${\rm U}(1)_{\rm X}$.\footnote{Of course one may suggest to introduce a scalar charged under ${\rm U}(1)_{\rm X}$ with a vanishing vacuum expectation value and a parabolic potential. However, again the natural mass scale for such a scalar is the UV cut-off of the theory. Since DM2 should be in thermal equilibrium with DR within a dark thermal bath, such a heavy scalar would be integrated-out easily when $T_{\rm DS}\simeq m_{\phi}$ holds, leaving no DM2 candidate in the scenario in the low energy regime. On the other hand, apparently we cannot take the gauge boson of ${\rm U}(1)_{\rm X}$ as the massive DM2 candidate since ${\rm U}(1)_{\rm X}$ should be unbroken.} Then it seems that we have again two choices: DM2 as a vector-like fermion and DM2 as a chiral fermion. Among these two choices, we quickly realized that the former must be the case because making chiral fermion DM2 massive requires a scalar charged under ${\rm U}(1)_{\rm X}$ inducing the spontaneous breaking of ${\rm U}(1)_{\rm X}$. Therefore, we cannot help but considering the vector-like fermion DM2. Note that the natural mass scale for such a vector-like fermion DM2 is the UV cut-off of the theory. Now we encounter a theoretically challenging, but very intriguing question: how can we make a mass of DM2 avoid being the UV cut-off of the theory?

In order to introduce a logically reasonable mass of the vector-like fermion DM2 different from the UV cut-off of the theory, we give our attention to the possibility where the mass is dynamically generated relying on a hidden strong dynamics. In general, for an asymptotically free ${\rm SU(N+4)}$ gauge theory with N fermions in the anti-fundamental representation ($\overline{\ytableausetup{textmode, centertableaux, boxsize=0.6em}
\begin{ytableau}
 \\
\end{ytableau}} $) and a single fermion in anti-symmetric representation (\ytableausetup{textmode, centertableaux, boxsize=0.6em}
\begin{ytableau}
 \\
 \\
\end{ytableau} ), the spectrum of the theory contains N(N+1)/2 composite massless baryonic bound states in an energy scale below the confinement~\cite{Dimopoulos:1980hn}. The theory enjoys the unbroken ${\rm SU(N)}_{f}\otimes {\rm U}(1)_{f}$ flavor symmetry and the N(N+1)/2 composite bound states correspond to the symmetric representation of ${\rm SU(N)}_{f}$. Requirement of vanishing ${\rm U}(1)_{f}\times[{\rm SU(N+4)}]^{2}$ anomaly provides us with $-(N+2)/N$ as the ratio of ${\rm U}(1)_{f}$ charges of $\overline{\ytableausetup{textmode, centertableaux, boxsize=0.6em}
\begin{ytableau}
 \\
\end{ytableau}}$ to \ytableausetup{textmode, centertableaux, boxsize=0.6em}
\begin{ytableau}
 \\
 \\
\end{ytableau} . With this, 't Hooft anomaly consistency conditions~\cite{tHooft:1979rat} were checked for three types of anomalies of $[{\rm U}(1)_{f}]^{3}$, $[{\rm SU(N)}_{f}]^{3}$ and ${\rm U}(1)_{f}\times[{\rm SU(N)}_{f}]^{2}$ to confirm the presence of N(N+1)/2 composite bound states~\cite{Dimopoulos:1980hn}.

Now applying the above observation to our case, we consider the simplest N=1 case to have a hidden ${\rm SU}(5)_{\rm X}$ gauge theory accompanied by a single fermion $\chi_{\alpha}$ in $\textbf{5}^{*}$ representation and the other single fermion $\xi_{\alpha\beta}$ in $\textbf{10}$ representation ($\alpha,\beta$: ${\rm SU}(5)_{\rm X}$ group indices). We may promote the global ${\rm U}(1)_{f}$ to the gauged ${\rm U}(1)_{\rm X}$ with the five chiral fermions given in Eq.~(\ref{eq:fivechiralfermions}) kept neutral to ${\rm SU}(5)_{\rm X}$. This implies that the ${\rm U}(1)_{\rm X}$ charge ratio of $\chi_{\alpha}$ and $\xi_{\alpha\beta}$ amounts to $-3$. We can take $Q_{\chi}=3$ and $Q_{\xi}=-1$. Now since ${\rm U}(1)_{\rm X}$ is the gauge symmetry, the theory needs to ensure that $[{\rm U}(1)_{\rm X}]^{3}$ and ${\rm U}(1)_{\rm X}\times[{\rm gravity}]^{2}$ should be anomaly-free for both energy regimes higher and lower than the confinement scale of  ${\rm SU}(5)_{\rm X}$. These two anomalies are already zero for five chiral fermions in Eq.~(\ref{eq:fivechiralfermions}). Thus we only need to check the conditions by taking into account contributions from $\chi_{\alpha}$ and $\xi_{\alpha\beta}$. As one can quickly check, the addition of one more chiral fermion $\psi'_{-5}$ neutral to ${\rm SU}(5)_{\rm X}$ helps the two anomalies vanish for $(\chi_{\alpha},\xi_{\alpha\beta},\psi'_{-5})$ provided the ${\rm U}(1)_{\rm X}$ charge of $\psi'_{-5}$ is $-5$ (the subscript stands for $Q_{\psi^{'}}=-5$).

As promised, now this set-up indeed yields very interesting mass term for the vector-like fermion DM2. In the energy regime lower than the confinement scale of ${\rm SU}(5)_{\rm X}$, say $\Lambda_{\rm X}$, the anomaly matching condition predicts the presence of the single massless composite baryonic state defined to be
\beq
\Omega_{+5}\equiv\frac{\chi_{\alpha}\xi_{\alpha\beta}\chi_{\beta}}{\Lambda_{\rm X}^{3}}\quad,\quad(Q_{\Omega}=+5)\,.
\label{eq:Omega}
\eeq
As the ${\rm SU}(5)_{\rm X}$ singlet bound state, we see that ${\rm U}(1)_{\rm X}$ charge of the bound state is +5 (and hence $+5$ in the subscript). Therefore, together with the chiral fermion $\psi'_{-5}$, it forms the following mass term for a vector-like fermion DM2\footnote{Now we have two chiral fermions with U(1)$_{\rm X}$ charge $-5$. We call a linear combination of them coupling to $\Omega_{+5}$ $\psi'_{-5}$ and the other orthogonal direction $\psi_{-5}$.}
\beq
\mathcal{L}\supset c\frac{\Lambda_{\rm X}^{3}}{M_{P}^{2}}\Omega_{+5}\psi'_{-5}+{\rm h.c.}=m_{\rm DM2}\Omega_{+5}\psi'_{-5}+{\rm h.c.}\,,
\label{eq:composite}
\eeq
where $c$ is a dimensionless coefficient. For the second equality, we defined the mass parameter for DM2 candidate to be
\beq
m_{\rm DM2}=c\Lambda_{\rm X}^{3}/M_{P}^{2}\,.
\label{eq:mDM2}
\eeq
This way of producing a vector-like fermion mass term dynamically was invoked in Refs.~\cite{ArkaniHamed:1998pf,Kamada:2019gpp,Kamada:2019jch} as well.

We conclude this section by summarizing the DS model built so far based on the requisites to resolve $H_{0}$ and $\sigma_{8}$ tensions. The DS we proposed entails ${\rm SU}(5)_{\rm X}\otimes {\rm U}(1)_{\rm X}$ gauge symmetry group under which the Standard model (SM) particle contents are singlet. The particle contents of the DS consist of
\begin{enumerate}
    \item ${\rm SU}(5)_{\rm X}$ singlet five chiral fermions given in Eq.~(\ref{eq:fivechiralfermions}) charged under ${\rm U}(1)_{\rm X}$
    \item $\chi_{\alpha}$ and $\xi_{\alpha\beta}$ transforming as \textbf{5}$^{*}$ and \textbf{10} representations of ${\rm SU}(5)_{\rm X}$ with ${\rm U}(1)_{\rm X}$ charges $Q_{\chi}=3$ and $Q_{\xi}=-1$ respectively
    \item ${\rm SU}(5)_{\rm X}$ singlet chiral fermion $\psi'_{-5}$ with ${\rm U}(1)_{\rm X}$ charge $Q_{\psi^{'}}=-5$
\end{enumerate}

\begin{table}[t]
\centering
\begin{tabular}{|c||c|c|} \hline
 & ${\rm SU}(5)_{\rm X}$ & ${\rm U}(1)_{\rm X}$ \\
\hline\hline
$\psi_{-9}$      &  1  &  $-9$  \\
$\psi_{-5}$  &  1  &  $-5$ \\
$\psi_{-1}$ &  1  &  $-1$  \\
$\psi_{+7}$      &  1  &  $+7$  \\
$\psi_{+8}$      & 1 &  $+8$   \\
$\psi'_{-5}$ &  1  &  $-5$ \\
$\chi_{\alpha}$ & \textbf{5}$^{*}$ &  $+3$  \\
$\xi_{\alpha\beta}$      &  \textbf{10} & $-1$ \\
\hline
\end{tabular}
\caption{Quantum numbers of the particle contents of the model. The shown are the particle contents for energy scales greater than ${\rm SU}(5)_{\rm X}$ confinement scale ($\Lambda_{\rm X}$). Below the $\Lambda_{\rm X}$, two $\chi_{\alpha}$ and a single $\xi_{\alpha\beta}$ form the bound state $\Omega_{\rm X}$ in Eq.~(\ref{eq:Omega}).}
\label{table:qn} 
\end{table}
This is summarized in Table.~\ref{table:qn}. Note that the quantum numbers of the set $(\chi_{\alpha}, \xi_{\alpha\beta},\psi'_{-5})$ under ${\rm SU}(5)_{\rm X}\otimes {\rm U}(1)_{\rm X}$ is the interesting reminiscent of the SM quarks, leptons and the right handed neutrino with the gauge group ${\rm SU}(5)_{\rm GUT}\otimes {\rm U}(1)_{\rm B-L}$.\footnote{The similar gauge structure between the SM and the dark sector was considered in \cite{Kamada:2019gpp,Kamada:2019jch} in the context of the unification between the two sectors and a strongly interacting DM.}

In the energy regime lower than the confinement scale ($\Lambda_{\rm X}$), two $\chi_{\alpha}$s and a single $\xi_{\alpha\beta}$ form the massless bound state $\Omega_{+5}$ given in Eq.~(\ref{eq:Omega}). Along with $\psi'_{-5}$, this bound state $\Omega_{+5}$ form the Dirac fermion $\Psi_{\rm DM2}\equiv(\psi'_{-5},\Omega_{+5}^{*})^{\rm T}$ with the mass parameter $m_{\rm DM2}$. The massless ${\rm U}(1)_{\rm X}$ gauge boson and five chiral fermions ($\psi$s) serve as DR, and $\Psi_{\rm DM2}$ is taken to be the candidate of the sub-dominant component of DM. ${\rm U}(1)_{\rm X}$ gauge interaction explains DR-DR and DM2-DR interactions. We assume a negligibly small (or zero) kinetic mixing between the gauge boson ($A_{\mu}'$) of ${\rm U}(1)_{\rm X}$ and that ($B_{\mu}$) of the hypercharge ${\rm U}(1)_{\rm Y}$ gauge symmetry of the SM.\footnote{As will be discussed in the next section, large massless degrees of freedom in the DS in our model demands a temperature of the DS lower than that of the SM sector. For that reason, if non-vanishing, as the only portal for the DS to rely on for communicating with the SM sector, the kinetic mixing must be suppressed enough for preventing thermal equilibrium between the DS and the SM sector. Thus for consistency it is not unnatural to assume a negligibly small kinetic mixing.}


\section{Constraining the Model}
\label{sec:parameter}
In this section, we discuss a parameter space of the model for resolving $H_{0}$ and $\sigma_{8}$ tensions. For alleviating the Hubble tension, as was discussed in Sec.~\ref{sec:InteractingDR}, we take the exemplary value of $\Delta N_{\rm fld}^{\rm CMB}\simeq0.4$ contributed by the interacting five chiral fermions shown in Eq.~(\ref{eq:fivechiralfermions}) and the massless dark photon $A'_{\mu}$. This would constrain the DS temperature as we shall see in Sec.~\ref{sec:DR}. As a resolution to the $\sigma_{8}$ tension, requiring (1) DM2-DR interaction to be efficient until $\rho_{\rm DR}$ becomes negligible in comparison with $\rho_{\rm DM2}$ after recombination and (2) $\Psi_{\rm DM2}$ to account for $2\%$ of the current DM relic abundance would constrain the gauge coupling of ${\rm U}(1)_{\rm X}$ and the mass of DM2. We discuss the resulting $(m_{\rm DM2},g_{\rm X})$ space in Sec.~\ref{sec:DM}.


\subsection{Dark Radiation}
\label{sec:DR}
Since we assume a negligibly small kinetic mixing between $A_{\mu}'$ and $B_{\mu}$, the model is in need of a way of producing the DS entities without the aid of the SM thermal bath. Among many, as an example, we focus on the possibility where a gauge singlet inflaton scalar field ($\Phi_{\rm I}$) produces dark sector particles via its decay. For simplicity of the scenario, we take $T_{\rm RH}<\Lambda_{\rm X}$ so as to have ${\rm SU}(5)_{\rm X}$ in the confined phase at the reheating era. 

Since every elementary fermion in the dark sector is chiral, the lowest dimension allowed operator coupling $\Phi_{\rm I}$ to the DS fermions is 
\beq
\mathcal{L}\supset c_{\Phi_{\rm I} f}\frac{\Phi_{\rm I}}{M_{P}}f^{\dagger}\bar\sigma^\mu D_\mu f\,,
\label{eq:inflatondecay}
\eeq
where $f$ denotes chiral fermions given in Table.~\ref{table:qn} and $c_{\Phi_{\rm I} f}$ is a dimensionless coefficient. Since all chiral fermions are massless,  we see that $\Phi_{\rm I}$ decays to $f$, $f^{\dagger}$ and an associated gauge boson. Since $g_{5\rm X}>g_{\rm X}$ at the reheating time where $g_{5\rm X}$ is the ${\rm SU}(5)_{\rm X}$ gauge coupling, we expect that the main decay mode of $\Phi_{\rm I}$ to the dark sector is $\Phi_{\rm I}\rightarrow\chi(\xi)+\chi^{\dagger}(\xi^{\dagger})+G'_{\mu}$ where $G'_{\mu}$ is the ${\rm SU}(5)_{\rm X}$ gauge boson. The rate of the decay of $\Phi_{\rm I}$ to a pair of ${\rm SU}(5)_{\rm X}$ charged fermions and $G'_{\mu}$ reads
\beq
\Gamma(\Phi_{\rm I}\rightarrow\chi(\xi)+\chi^{\dagger}(\xi^{\dagger})+G'_{\mu})\sim\kappa\frac{m_{\Phi_{\rm I}}^{3}}{M_{P}^{2}}\,,
\label{eq:decayrate}
\eeq
where $m_{\Phi_{\rm I}}$ is the inflaton mass and $\kappa$ is $\mathcal{O}(10^{-2})-\mathcal{O}(1)$ dimensionless coefficient.\footnote{From the phase space integral of the three-body decay, the decay rate is accompanied by the factor of $(2^{6}\pi^{3})^{-1}$. On production, however, the fermions and $G'_{\mu}$ will quickly go through hadronization, converting the energy of $\Phi_{\rm I}$ to hadrons. Taking into account these, we make a rough estimate for $\kappa$. The precise value of $\kappa$ is not important for our purpose.} We anticipate that the scattering of a subset of $\chi$s and $\xi$s produces $A'_{\mu}$ and chiral fermion DRs in Eq.~(\ref{eq:fivechiralfermions}) to form a dark thermal bath once $\Gamma({\rm DM2+DM2}\rightarrow{\rm DR+DR})\simeq g_{\rm X}^{4}T_{\rm DS}>H$ is satisfied. We notice that DR self-interaction is turned on at the time of $g_{\rm X}^{4}T_{\rm DS}>H$ and thus the active DR-DR interaction becomes persistent since the formation of the dark thermal bath. The ratio of the energy density of the DS to that of the inflaton is given by
\beq
{\rm Br}\equiv\frac{\rho_{\rm DS}}{\rho_{\rm \Phi_{\rm I}}}\simeq\frac{\rho_{\rm DS}}{\rho_{\rm SM}}\simeq\frac{\Gamma(\Phi_{\rm I}\rightarrow\chi(\xi)+\chi^{\dagger}(\xi^{\dagger})+G'_{\mu})}{\Gamma(\Phi_{\rm I}\rightarrow {\rm SM\,\,particles})}\,.
\label{eq:Br}
\eeq
For the second relation, we used the approximation $\rho_{\rm SM}\simeq\rho_{\rm \Phi_{\rm I}}$, assuming the most of the energy of $\Phi_{\rm I}$ is used for creating the SM particles. The degree of freedom of radiations for the DS at the reheating time is $g_{*,{\rm DR}}(a_{\rm RH})=2+(7/8)\times(5\times2+4)=14.25$ while for the SM we have $g_{*,{\rm SM}}(a_{\rm RH})=106.75$. Here $a_{\rm RH}$ is a scale factor for the reheating time. $g_{*,{\rm DR}}(a_{\rm RH})$ is contributed by $A'_{\mu}$, DM2, and five chiral fermions in Eq.~(\ref{eq:fivechiralfermions}). 

At the recombination era, the amount of the extra relativistic species reads
\beqs
\Delta N_{\rm fld}^{\rm CMB}&\simeq&\frac{8}{7}\left(\frac{11}{4}\right)^{4/3}\frac{\rho_{{\rm DR}}}{\rho_{\gamma}}\cr\cr
&=&\frac{8}{7}\left(\frac{11}{4}\right)^{4/3}\left(\frac{g_{*,{\rm DR}}(a_{\rm rec})}{g_{*,\gamma}(a_{\rm rec})}\right)\left(\frac{T_{\rm DR}(a_{\rm rec})}{T_{\rm SM}(a_{\rm rec})}\right)^{4}\,.
\label{eq:Neff}
\eeqs
Using $\Delta N_{\rm fld}^{\rm CMB}=0.4$, $g_{*,{\rm DR}}(a_{\rm rec})=2+(7/8)\times(5\times2)=10.75$ and $g_{*,\gamma}(a_{\rm rec})=2$, we obtain $T_{\rm DR}(a_{\rm rec})/T_{\rm SM}(a_{\rm rec})\simeq0.36$. For our purpose of estimating an order of magnitude for Br in Eq.~(\ref{eq:Br}), it suffices to identify $T_{\rm DR}(a_{\rm RH})/T_{\rm SM}(a_{\rm RH})$ with $T_{\rm DR}(a_{\rm rec})/T_{\rm SM}(a_{\rm rec})$ based on the approximation that $T_{\rm SM},T_{\rm DR}\propto1/a$. Thus, from Eq.~(\ref{eq:Br}), $g_{*,{\rm DR}}(a_{\rm RH})=14.25$ and $g_{*,{\rm SM}}(a_{\rm RH})=106.75$, we obtain ${\rm Br}\simeq(14.25/106.75)\times0.36^{4}\simeq2\times10^{-3}$. 

On the other hand, at the reheating time the expansion rate of the universe is expected to be comparable to the rate of the inflaton decay to SM particle for reheating, i.e. $\Gamma(\Phi_{\rm I}\rightarrow {\rm SM\,\,particles})\simeq H(a=a_{\rm RH})\simeq T_{\rm RH}^{2}/M_{P}$ where $T_{\rm RH}=T_{\rm SM}(a_{\rm RH})$ holds and $M_{P}\simeq2.4\times10^{18}{\rm GeV}$ is the reduced Planck mass. Using this relation, Eq.~(\ref{eq:decayrate}) and Eq.~(\ref{eq:Br}) with ${\rm Br}\simeq2\times10^{-3}$, we obtain
\beq
T_{\rm RH}\simeq\sqrt{\kappa}\times10^{7}\times(\frac{m_{\Phi_{\rm I}}}{10^{10}{\rm GeV}})^{3/2}{\rm GeV}\,.
\label{eq:TRH}
\eeq
Practically, in order that the inflaton $\Phi_{\rm I}$ can produce ${\rm SU}(5)_{\rm X}$-charged fermions with the dynamical mass of the order $\Lambda_{\rm X}$, $m_{\Phi_{\rm I}}$ should satisfy $m_{\Phi_{\rm I}}\gtrsim\Lambda_{\rm X}$. Given $\Lambda_{\rm X}\simeq10^{12}-10^{13}{\rm GeV}$, we realize that $m_{\Phi_{\rm I}}$ should be at least greater than $10^{13}{\rm GeV}$. Along this line of reasoning, Eq.~(\ref{eq:TRH}) is converted into
\beq
T_{\rm RH}\gtrsim\sqrt{\kappa}\times3\times10^{11}{\rm GeV}\,.
\eeq


\subsection{Dark Matter}
\label{sec:DM}
As $T_{\rm DS}$ becomes comparable to $m_{\rm DM2}$, the freeze-out of DM2 annihilation process to produce DRs will get started. In our model, there are two annihilation channels which are $t$-channel annihilation ${\rm DM2+DM2^{*}}\rightarrow A'_{\mu}+A'_{\mu}$ and $s$-channel annihilation ${\rm DM2+DM2^{*}}\rightarrow\psi_{Q}+\psi^{\dagger}_{Q}$ with $\psi_{Q}$ given in Eq.~(\ref{eq:fivechiralfermions}). The thermally averaged cross sections for these processes are given by
\begin{numcases}{<\!\!\sigma v\!\!>=}
  \frac{Q_{\rm DM}^{4}g_{\rm X}^{4}}{16\pi m_{\rm DM2}^{2}}, & $t$-channel\\
  \frac{Q_{\rm DM}^{2}Q_{\psi}^{2}g_{\rm X}^{4}}{32\pi m_{\rm DM2}^{2}}, &  $s$-channel
\label{eq:crosssection}  
\end{numcases}
Here $Q_{\rm DM}=-5$ and $Q_{\psi}=-9,-5,-1,+7,+8$. Adding all the contributions, we obtain
\beqs
<\!\!\sigma v\!\!>_{\rm total}&=&\frac{Q_{\rm DM}^{2}g_{\rm X}^{4}}{16\pi m_{\rm DM2}^{2}}\left(Q_{\rm DM}^{2}+\sum_{i=1}^{5}\frac{Q_{\psi_{i}}^{2}}{2}\right)\cr\cr&\simeq&67\times\frac{g_{\rm X}^{4}}{m_{\rm DM2}^{2}}\,.
\eeqs

\begin{figure}[t]
\centering
\hspace*{-5mm}
\includegraphics[width=0.45\textwidth]{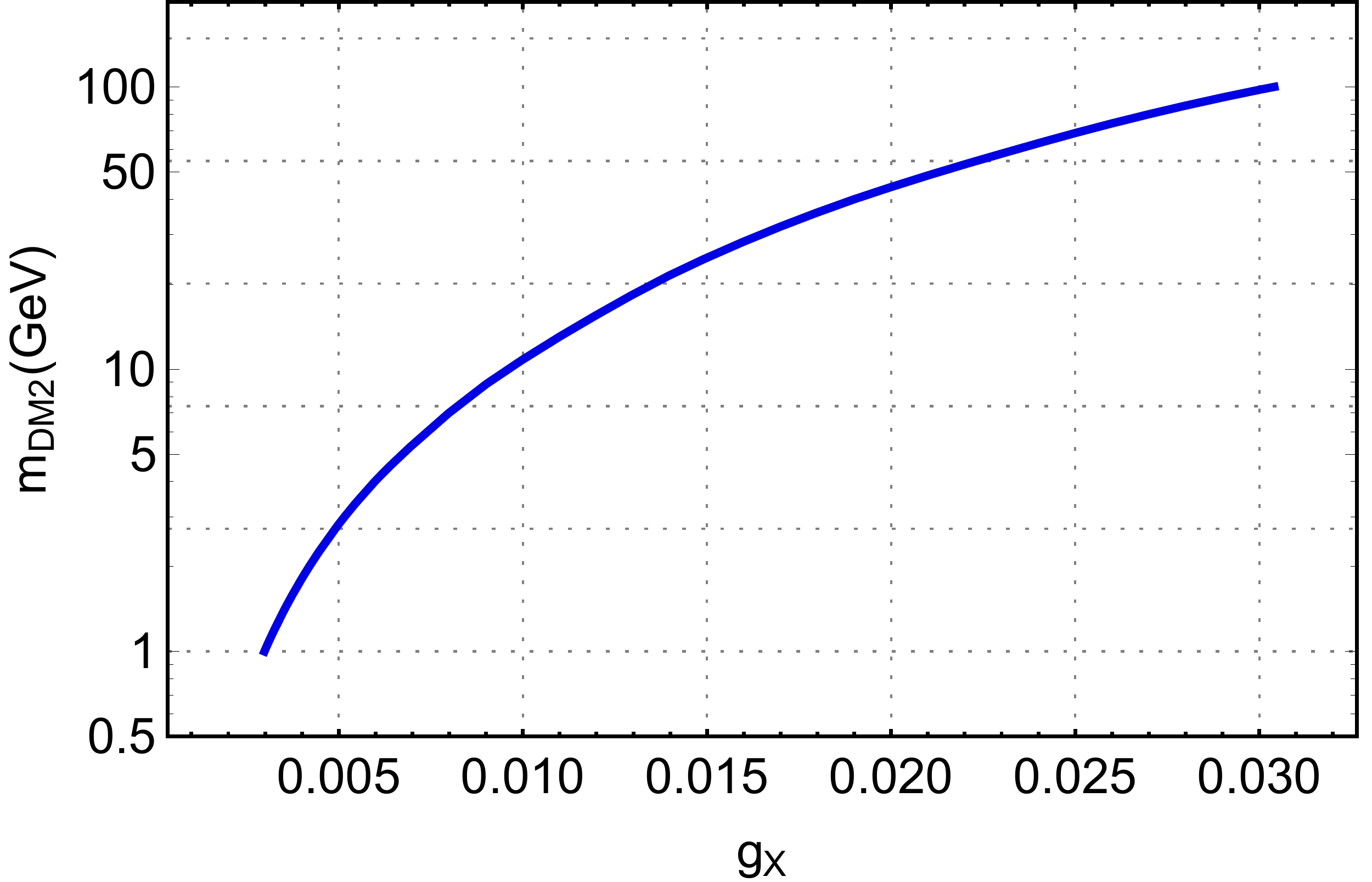}
\caption{The set of points in $(g_{\rm X},m_{\rm DM2})$ plane satisfying the relation $\Omega_{\rm DM2}=0.02\times\Omega_{\rm DM}$.}
\vspace*{-1.5mm}
\label{fig1}
\end{figure}

The current DM2 abundance (2\% of the DM population) reads
\beqs
\Omega_{\rm DM2,0}&=&\frac{\rho_{\rm DM2,0}}{\rho_{\rm cr,0}}=\frac{s_{0}}{\rho_{\rm cr,0}}\times m_{\rm DM2}\times Y_{\rm DM2}\simeq5\times10^{-3}\cr\cr&\simeq&1.35\times10^{10}\times\left(\frac{m_{\rm DM2}}{100{\rm GeV}}\right) \times Y_{\rm DM2}\,,
\label{eq:DMdensity}
\eeqs
where $Y_{\rm DM2}=n_{\rm DM2}/s$ is the comoving number density of DM2, and we used $\rho_{\rm cr,0}=8.02764\times10^{-11}\times h^{2}\,{\rm eV}^{4}$ ($h=0.7$) and $s_{0}=2.21\times10^{-11}{\rm eV}^{3}$ for the second line. $Y_{\rm DM2}$ is a function of $g_{\rm X}$ and $m_{\rm DM2}$ via its dependence on the thermally averaged cross section $<\!\!\sigma v\!\!>$. Given $T_{\rm DS}/T_{\rm SM}\simeq0.36$, for computation of $Y_{\rm DM2}$, we refer to Ref.~\cite{Chacko:2015noa}. We focus on the mass regime $m_{\rm DM2}\in(1{\rm GeV},100{\rm GeV})$ where the effect of Sommerfeld enhancement on the thermal DM relic abundance is negligible~\cite{Tulin:2013teo}. For this mass regime, we expect $\Lambda_{\rm X}\simeq10^{12}-10^{13}{\rm GeV}$ with $\mathcal{O}(1)$ coefficient $c$ for the operator shown in Eq.~(\ref{eq:composite}).
We search for points in the space $(g_{\rm X},m_{\rm DM2})$
obeying Eq.~(\ref{eq:DMdensity}) and show the result in Fig.~\ref{fig1}. 

In order for DM2-DR interaction to cause the desired suppression of the matter fluctuation at the scale $k\simeq0.2h{\rm Mpc}^{-1}$, it should remain efficient until the time when $3\rho_{\rm DM2}/(4\rho_{\rm DR})>\!\!>1$ is satisfied (slightly after recombination era). To this end, the momentum transfer rate corresponding to $(g_{\rm X},m_{\rm DM2})$ on the blue line in Fig.~\ref{fig1} should be greater than the Hubble expansion rate until the late times. The momentum transfer rate reads~\cite{Feng:2010zp}
\beqs
\Gamma_{\rm DM2-DR}&\sim& n_{\rm DR}<\!\!\sigma v\!\!>_{\rm DM2-DR}\frac{T_{\rm DS}}{m_{\rm DM}}\cr\cr&\simeq& Q_{\rm DM2}^{2}Q_{\rm DR}^{2}g_{\rm X}^{4}\frac{T_{\rm DS}^{2}}{m_{\rm DM2}}\,,
\label{eq:momentumtransfer}
\eeqs
where $n_{\rm DR}$ is the number density of DR and $<\!\!\sigma v\!\!>_{\rm DM2-DR}$ is the thermally averaged cross section for the $t$-channel DM2-DR scattering. Using $H(a)=H_{0}\sqrt{\Omega_{\rm rad,0}/a^{4}+\Omega_{\rm m,0}/a^{3}+\Omega_{\Lambda,0}}$ with $\Omega_{j}$ the relic abundance of the species $j$ and $H_{0}$ the Hubble expansion rate today, the numerical comparison of $\Gamma_{\rm DM-DR}$ in Eq.~(\ref{eq:momentumtransfer}) and $H(a)$ for $a>a_{\rm rec}$ can be done. The comparison shows indeed that the interaction between DM2 and DR are kept efficient after recombination by satisfying $\Gamma_{\rm DM-DR}>H$.


\section{Conclusion}
\label{sec:conclusion}
In this letter, we proposed a model of a dark sector which can address cosmological tensions for the expansion rate of the universe ($H_{0}$ tension) and matter fluctuations at $8h^{-1}{\rm Mpc}$ ($\sigma_{8}$ tension). The particle contents of the model include a dominant DM component (DM1), a subdominant DM component (DM2) and DR. And DM1 can be any kind of CDM. The model extends the SM by the additional unbroken gauge group ${\rm SU(5)_{\rm X}}\otimes{\rm U}(1)_{\rm X}$ and several massless chiral fermions making the theory anomaly free. In the UV regime near the Planck scale, beginning with  ${\rm SU(5)_{\rm X}}$ gauge coupling $g_{\rm 5X}\simeq0.5-0.6$, the ${\rm SU(5)_{\rm X}}$ gauge theory gets into the confined phase at $\Lambda_{\rm X}\simeq10^{12}-10^{13}{\rm GeV}$. As a consequence, in the low energy regime ($\mu<\Lambda_{\rm X}$),  there arises a Dirac fermion DM2 candidate with $m_{\rm DM2}=\mathcal{O}(1)-\mathcal{O}(100){\rm GeV}$ which interacts with DR candidate by exchanging ${\rm U}(1)_{\rm X}$ gauge boson. This DM2-DR interaction suppresses the growth of the density fluctuation of DM2 until the recombination era, resolving the $\sigma_{8}$ tension. ${\rm U}(1)_{\rm X}$ gauge coupling $g_{\rm X}=\mathcal{O}(10^{-3})-\mathcal{O}(10^{-2})$ enables DM2 with $m_{\rm DM2}=\mathcal{O}(1)-\mathcal{O}(100){\rm GeV}$ to account for $2\%$ of the current DM relic abundance. This strength of $g_{\rm X}$ is large enough to keep DM2-DR interaction efficient until late times after the recombination. Moreover, ${\rm U}(1)_{\rm X}$ gauge interaction introduces interaction among DRs that helps alleviate $H_{0}$ tension. 

The merit of the model lies in the fact that it can explain vector-like fermion DM2 mass based on the hidden strong dynamics without relying on a scalar condensation nor introducing a mass term by hand. In the absence of introduction of any new scalar, the model can avoid suffering from a new hierarchy problem. In addition, because DM2 is responsible for only few fraction of the total DM population, $g_{\rm X}=\mathcal{O}(10^{-2})$ required for matching relic abundance of GeV-scale DM2 does not lead to too much suppression of the matter power spectrum inconsistent with observation. With $\alpha_{\rm X}=g_{\rm X}^{2}/4\pi=\mathcal{O}(10^{-5})$ and $\rho_{\rm DM2}/\rho_{\rm DM}=0.02$, we see that the constraint on $(\alpha_{\rm X},m_{\rm DM})$ for a darkly charged DM given in Refs.~\cite{Feng:2009mn,Agrawal:2016quu} does not apply to ours for exclusion.


\begin{acknowledgments}
N. Y. is supported by JSPS KAKENHI Grant Number JP16H06492. T. T. Y. is supported in part by the China Grant for Talent Scientific Start-Up Project and the JSPS Grant-in-Aid for Scientific Research No. 16H02176, No. 17H02878, and No. 19H05810 and by World Premier International Research Center Initiative (WPI Initiative), MEXT, Japan. 

\end{acknowledgments}


\bibliography{main}

\end{document}